\documentclass[twocolumn, showpacs,preprintnumbers,amsmath,amssymb]{revtex4}
\usepackage{amssymb}
\usepackage{xcolor}
\usepackage{graphicx}
\usepackage{dcolumn}
\usepackage{bm}
\usepackage{multirow}
\usepackage{booktabs}
\usepackage{natbib}
\usepackage{soul}
\usepackage{amssymb}
\usepackage{amsmath}	
\usepackage{color}
\usepackage{cases}

\begin{document}
\preprint{APS/123-QED}

\title{Theoretical study of the excited states of NeH and of their non-adiabiatic couplings: a preliminary for the modeling of the dissociative recombination of NeH$^{+}$}

\author{R. Hassaine$^{1}$}\email[]{riyad.hassaine@univ-lehavre.fr}
\author{D. Talbi$^{2}$}\email[]{dahbia.talbi@umontpellier.fr}
\author{R. P. Brady$^{3}$}\email[]{ryan.brady.17@ucl.ac.uk}
\author{J. Zs. Mezei$^{1,4}$}\email[]{mezei.zsolt@atomki.hu}
\author{J. Tennyson$^{1,3}$}\email[]{j.tennyson@ucl.ac.uk}
\author{Ioan F. Schneider$^{1,5}$}\email[]{ioan.schneider@univ-lehavre.fr}
\affiliation{$^{1}$LOMC CNRS-UMR6294, Universit\'e le Havre Normandie, F-76058 Le Havre, France}%
\affiliation{$^{2}$LUPM CNRS-UMR5299, Universit\'e de Montpellier, F-34095 Montpellier, France}%
\affiliation{$^{3}$Dept. of Physics and Astronomy, University College London, WC1E 6BT London, UK}%
\affiliation{$^{4}$HUN-REN Institute for Nuclear Research (ATOMKI), H-4001 Debrecen, Hungary}%
\affiliation{$^{5}$LAC-UMR9188, CNRS Universit\'e Paris-Saclay, F-91405 Orsay, France}%
\date{\today}

\begin{abstract}
Potential energy curves and matrix elements of radial non-adiabatic couplings of $^{2}\Sigma^{+}$ and $^{2}\Pi$ states of the NeH molecule are calculated using the electronic structure package MOLPRO, in view of the study of the reactive collisions between low-energy electrons and NeH$^{+}$.
\end{abstract}

\maketitle

\section{\label{sec:Introduction}Introduction}

Neon is envisaged as a coolant in the divertor of the International Thermonuclear Experimental Reactor (ITER) \cite{zajfman1996, mitchell2001, mitchell2005, ngassam2008}, since it can absorb energy via its excitation and ionization through reactions with rapid electrons. Dielectronic recombination of the formed ions, followed by electronic relaxation process:
\begin{equation}
\mbox{Ne}^{+}+e^{-} \rightarrow \mbox{Ne}^{**} \rightarrow \mbox{Ne}+h\nu,
\end{equation}

\noindent allows safe dissipation of heat along the walls through radiation. On and near the surface of the reactor, collisions between neon atoms with vibrationally excited H$_{2}$$^{+}$ can produce NeH$^{+}$:
\begin{equation}
\mbox{Ne}+\mbox{H}_{2}^{+}(v_{i}^{+}>1) \rightarrow \mbox{NeH}^{+}+\mbox{H},
\end{equation}

\noindent the reaction for ground state ions being endothermic by 0.54 eV.\cite{mitchell2005} 
The presence of NeH$^{+}$ in the edge plasma implies the need for quantifying the reactivity of this ion. 
This can be achieved from the calculation of the cross sections and rate coefficients which characterize the efficiency of collisions between NeH$^{+}$ and other species present in the plasma. Among them, 
the collisions with electrons, resulting in
 dissociative recombination (DR)\cite{ngassam2008} and vibrational transitions (VT):
\begin{equation}
\mbox{NeH}^{+}+e^{-} \rightarrow \mbox{NeH}^{*},\mbox{NeH}^{**} \rightarrow
\left\{
    \begin{array}{lll}
         \mbox{Ne}+\mbox{H} \quad\quad\,\,\, (\text{DR})\\
        \mbox{NeH}^{+*}+e^{-} \,\, (\text{VT})\\
    \end{array}
    ,
\right.
\end{equation}

\noindent
are particularly important.

Cross sections for the DR of NeH$^{+}$ were already measured at the storage ASTRID ring two decades ago~\cite{mitchell2005}. Theoretical evaluations of the cross sections have also been performed for collision energies ranging between 5 and 22 eV~\cite{ngassam2008}, and they agree with the experimental ones between 6 and 10 eV. So far, theoretical results are not available for the physically important electron collision energies below 5 eV, where experiment~\cite{mitchell2005} measures significant cross sections of about 10$^{-18}$ cm$^{2}$.

The low energy DR of the hydride cations relying on the elements preceding and following neon in the periodic tables - helium and argon respectively - is well understood and quantified.
Both present in the edge plasma, they were object of detailed investigations related mainly on their abundances in the interstellar media.

HeH$^+$, one of the oldest molecule in the Universe, 
was only recently observed in the nebula NGC 7027~\cite{gusten2019}. It is the simplest molecular prototype that recombines with electrons exclusively through the indirect pathway, proceeding via electron capture into Rydberg states, without having a diabatic dissociative neutral state that crosses the cation~\cite{jt152,guberman1994}. The initial state-selective measurements performed on the Cryogenic Storage Ring~\cite{novotny2019} have shown a dramatic decrease of the DR rate coefficients at very low collision energies, that can cause higher abundances in cold interstellar environments. The most recent calculations of \v{C}urik et al~\cite{vcurik2020a, hvizdovs2020b} using energy-dependent rovibrational frame transformation combined with multichannel quantum defect theory have confirmed the experimental observations. 

 As for ArH$^+$, it was detected more than one decade ago in the Crab Nebula~\cite{barlow2013}.
Prior to its discovery, Mitchell et al~\cite{mitchell2005b},
working in the ASTRID storage ring, had estimated the anisotropic rate coefficient below 2 eV lower than 10$^{-9}$
cm$^3$s$^{-1}$. Some of us addressed the 
molecular structure of ArH~\cite{abdoulanziz2018} and found that the asymptotic limit of the lowest diabatic electronic state of this molecule is 1.822 eV above the ground vibrational level, which implies - if restricted exclusively to  Rydberg-valence coupling - no recombination below this energy and, consequently, 
a negligible rate coefficient~\cite{djuissi2022}. 
However, Kalosi et al~\cite{kalosi2024} measured low but non-negligible DR rates into the ground state of ArH, claiming the mechanism of non-adiabatic coupling between the ionization continuum and the ground electronic state. This scenario is supported by theoretical considerations in the same study, as well as computations in progress by Larson and Orel~\cite{kalosi2024}.

Unlike HeH$^{+}$~\cite{gusten2019} and ArH$^{+}$~\cite{barlow2013}, NeH$^{+}$ remains undetected in these media. Recent work by Sil et al\cite{sil2024} suggests that NeH\(^+\) could be formed in nova ejecta  via the reaction:
\[
\text{HeH}^+ + \text{Ne} \rightarrow \text{NeH}^+ + \text{He},
\]
with a rate coefficient of approximately \(10^{-9} \, \text{cm}^3 \, \text{s}^{-1}\). This pathway appears more likely than the alternative reaction proposed by Theis et al\cite{theis2015}:
\[
\text{Ne}^+ + \text{H}_2 \rightarrow \text{NeH}^+ + \text{H},
\]
which is considered unfavorable. The formation of NeH\(^+\) is expected to be heavily dependent on the abundance of HeH\(^+\), which is more prevalent due to the significant presence of helium and atomic hydrogen in the environment. However, NeH\(^+\) production is constrained by the comparatively low cosmological abundance of neon\cite{schwarz2002} and the underabundance of molecular hydrogen. Another potential explanation for the non-detection of NeH\(^+\) may lie in its dissociative recombination behavior. A high dissociative recombination rate coefficient at interstellar medium relevant temperatures could significantly reduce its steady-state concentration, making detection more challenging.

Below 5 eV, the lack of theoretical data for the cross sections can be initially addressed by producing the \textit{ab initio} adiabatic potential energy curves (PECs) of NeH corresponding to the bound Rydberg states energetically open for dissociation (below $v_{i}^{+}=0$ of the ion's ground state), similarly to Refs.~\cite{theodorakopoulos1984, Theodorakopoulos1987, baer1994, petsalakis1998, lo2005}, and non-adiabatic couplings responsible for the transitions among the different Born-Oppenheimer adiabatic states. Here we consider both the relevant NeH PECs and compute the matrix elements which give the radial non-adiabatic couplings (NACs) between all the explored states of the neutral. In the current study, we focus on these two aspects, while the dynamical calculations will be considered in a further work, relying on the current molecular structure data results.

The organisation of the paper is as follows: In section II, the computational steps are presented. Section III contains the results - PECs, spectroscopic data, NACs - and we provide the concluding remarks in section IV.

\section{Computational details}{\label{sec:computational}}

The potential energy curve calculations for NeH and NeH$^{+}$ in its ground state were carried out using the MOLPRO quantum chemistry program suite~\cite{werner2022} at the MCSCF-MRCI level of theory with a complete active space (CAS) of 8,3,3,0 orbitals in the C$_{2v}$ point-group symmetry. It corresponds to the complete valence active space extended to include the $n=2$ and $n=3$ orbitals of hydrogen. We started with the augmented correlation-consistent polarized valence triple zeta (aug-cc-pVTZ) basis sets implemented in MOLPRO, and we have extended the hydrogenic part of the basis by two $s$, three $p$, and one $d$ diffuse orbitals (AO) with exponents from Ref.~\cite{baer1994}, which are given in table~\ref{tab0}.

\begin{table}[htbp]
	\caption{H atom Rydberg basis set.}
	\label{tab0}
	\centering
	\begin{tabular}{cccccccc}
		\hline
	Type & Exponent & Coefficient \\
	    \hline
     \rule{0pt}{3ex}$s$ & 0.006685 & 1.0\\
     $s$ & 0.002670 & 1.0\\
     $p$ & 0.024684 & 1.0\\
     $p$ & 0.007169 & 1.0\\
     $p$ & 0.002867 & 1.0\\
     $d$ & 0.003600 & 1.0\\
     \hline
     	\end{tabular}
\end{table}

In this way, the chosen basis set provides a more accurate description of the states correlating to the $n=2$ and $n=3$ hydrogen dissociation limits.

For the neutral molecule, the MCSCF wave functions were optimized by state-averaging the five lowest $^{2}\Sigma^{+}$ states (5 A$_{1}$ states in the C$_{2v}$ symmetry of the calculation) and the two lowest $^{2}\Pi$ states ($^2$B$_{2}$/$^2$B$_{1}$ states in the C$_{2v}$ symmetry of the calculation). To summarize, the ground state of NeH$^{+}$, X $^{1}\Sigma^{+}$, and the lowest 7 states of NeH, respectively X $^{2}\Sigma^{+}$, A $^{2}\Sigma^{+}$, B $^{2}\Pi$, C $^{2}\Sigma^{+}$, 4 $^{2}\Sigma^{+}$, 2 $^{2}\Pi$, and 5 $^{2}\Sigma^{+}$ have been calculated.

The NACs between states of the neutral were calculated using the following expressions~\cite{guberman1994, brady2024}:

\begin{widetext}
\begin{equation}
A_{ij}(R)=\biggl \langle \psi^{\Gamma}_{i}(\{\Vec{r}\},R)\bigg|\frac{\partial}{\partial R}\bigg|\psi^{\Gamma}_{j}(\{\Vec{r}\},R)\biggr \rangle_{\{\Vec{r}\}}
\label{AR}
\end{equation}

\begin{equation}
B_{ij}(R)=\biggl \langle \psi^{\Gamma}_{i}(\{\Vec{r}\},R)\bigg|\frac{\partial^{2}}{\partial R^{2}}\bigg|\psi^{\Gamma}_{j}(\{\Vec{r}\},R)\biggr \rangle_{\{\Vec{r}\}}=\frac{\partial A_{ij}(R)}{\partial R} - A_{ij}^{2}(R)
\label{BR}
\end{equation}
\end{widetext}

\noindent Here $\Gamma$ stands for the symmetry of the molecular state, $i$ and $j$ denote different electronic states belonging to the same symmetry, and we focus exclusively on couplings that satisfy $i<j$. $R$ is the internuclear distance, \{$\vec{r}$\} denotes the complete set of electronic coordinates linked to the electrons of the neutral. In order to calculate the NACs using Eqs.~(\ref{AR}) and (\ref{BR}), one has to integrate over the electronic coordinates. This is achieved numerically using the MOLPRO derivative couplings (DDR) procedure for $A(R)$, and by using central finite difference scheme for the derivative to obtain $B(R)$. The spectroscopical description of the target, including the 15 vibrational levels of the ground state of NeH$^{+}$ were obtained by using the Numerov-Cooley method~\cite{noumerov1924, numerov1927} to solve the nuclear-motion Schr\"{o}dinger equation.

\section{Results and discussions}{\label{sec:results}}

\begin{figure}
\centering
		\includegraphics[width=0.95\columnwidth]{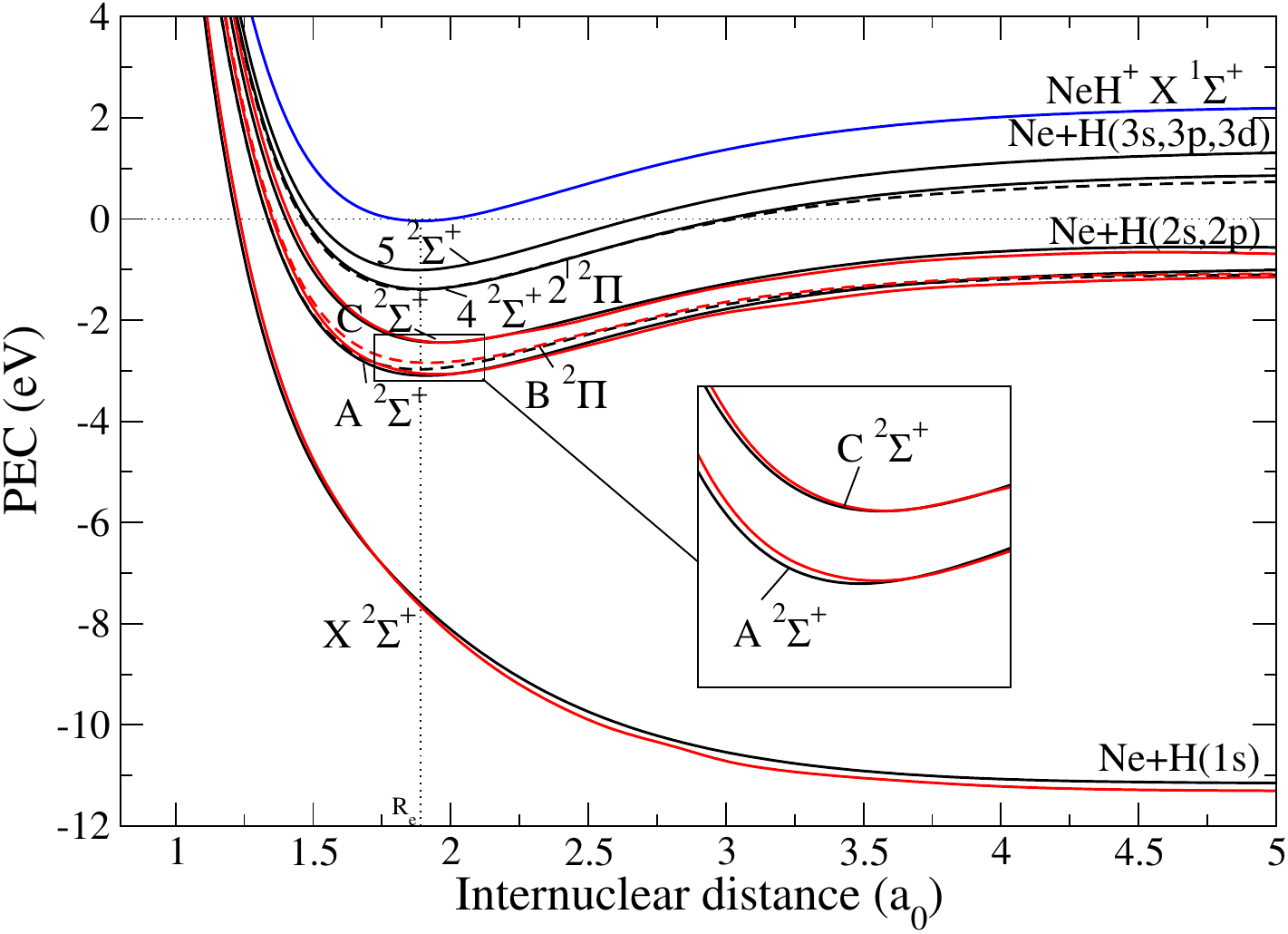}
    
\caption
{\textit{Ab initio} PECs of the ground electronic state of NeH$^{+}$ and of the lowest - ground and Rydberg - electronic states of NeH. Black solid lines stand for the neutral with $^{2}\Sigma^{+}$ symmetry, black dashed lines for the neutral with $^{2}\Pi$ symmetry, and the solid blue line corresponds to the ion's ground state X $^{1}\Sigma^{+}$. Our results are compared with the calculations of Theodorakopoulos et al.\cite{Theodorakopoulos1987} (red lines).}
    \label{fig:1}
\end{figure}

Figure~\ref{fig:1} displays the \textit{ab initio} PECs of the ground electronic state of NeH$^{+}$ ($^{1}\Sigma^{+}$) (blue curve), the repulsive ground electronic state of NeH, and the mono-excited Rydberg states of NeH for both symmetries $^{2}\Sigma^{+}$ and $^{2}\Pi$, up to the Ne+H($n=3$) dissociation limit. The dissociative ground state of the neutral X $^{2}\Sigma^{+}$ correlates to the Ne+H($1s$) atomic limit, where Ne stands for the $^1$S$_0$ ground state of atomic neon. 
The excited states correlating with the Ne+H($n=2$) limits from bottom to top are A $^{2}\Sigma^{+}$ - B $^{2}\Pi$ - C $^{2}\Sigma^{+}$, and the states tending to the Ne+H($n=3$) are 4 $^{2}\Sigma^{+}$ - 2 $^{2}\Pi$ - 5 $^{2}\Sigma^{+}$. Our results - black curves in Fig.~\ref{fig:1} - compare well with those from Theodorakopoulos et al.\cite{Theodorakopoulos1987} - red curves. 

\begin{figure}
\centering
		\includegraphics[width=0.95\columnwidth]{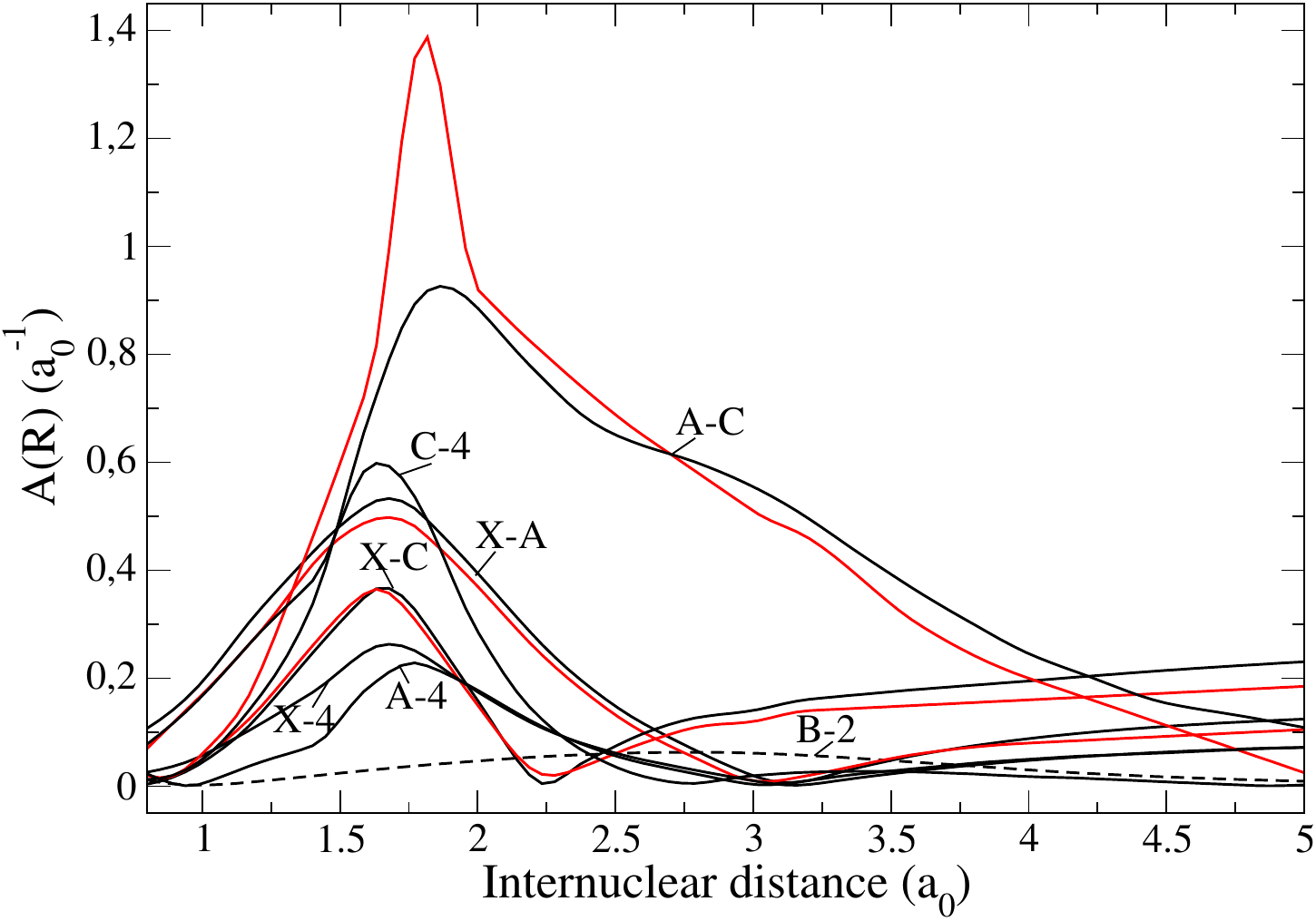}
\caption
{Radial non-adiabatic couplings $A(R)$. The color code is the same as for Fig.~\ref{fig:1}.}
    \label{fig:2}
\end{figure}

Figure~\ref{fig:2} shows the NACs $A(R)$ corresponding to Eq.~(\ref{AR})between the molecular states having $^{2}\Sigma^{+}$ (solid curves) and $^{2}\Pi$ (dashed curves) symmetry, up to the Ne+H($n=3$) dissociation limit. The couplings involving the highest excited state 5 $^{2}\Sigma^{+}$ are not shown as their magnitudes are negligible (approximately $10^{-8}$ $a_{0}^{-1}$). One can notice that the couplings of the $^{2}\Sigma^{+}$ states are the more important ones, while the coupling involving the two $^{2}\Pi$ states is about one order of magnitude smaller. Similarly to PECs, we have compared our NACs (in black) with those calculated by Theodorakopoulos et al~\cite{Theodorakopoulos1987} presented in red. One can conclude that the current results are consistent with the previous ones.  
The sharp peak of the A-C coupling is due to the strong interaction between these two electronic states.

\begin{figure}
\centering
		\includegraphics[width=0.95\columnwidth]{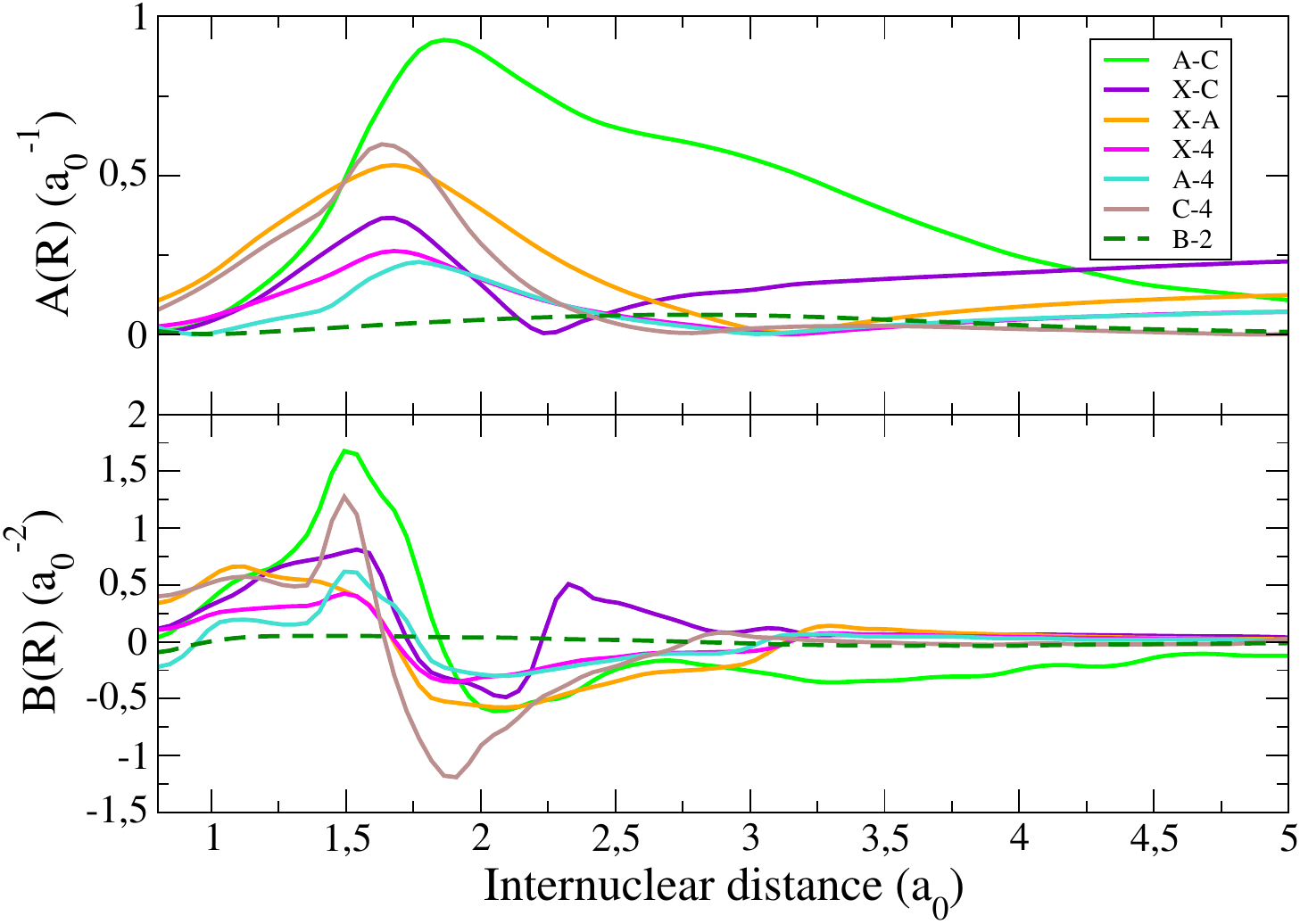}
    
\caption
{First-derivative radial non-adiabatic couplings $A(R)$ and second-derivative radial non-adiabatic couplings $B(R)$, between the states of NeH.}
    \label{fig:3}
\end{figure}

Figure~\ref{fig:3} contains all the NACs $A(R)$ given by Eq.~(\ref{AR})~(already shown in Fig.~\ref{fig:2}), in comparison with $B(R)$ given by Eq.~(\ref{BR}). One can see that the most significant  $B(R)$ NAC is the A-C one, followed by C-4, X-A, X-C, X-4, A-4, B-2, which predicts the importance of the Rydbergs correlating with the Ne+H($n=3$) limit for the inter-nuclear dynamics.

\begin{table}[t]
	\caption{
    Equilibrium separation ($R_{e}$), absolute minimum energy ($E(R_{e})$) and well-depth ($D_{e}$) for NeH$^{+}$ (X $^{1}\Sigma^{+}$) PECs:  comparison with theoretical as well as experimental data from the literature. $\epsilon_{0}$ represents percentage difference between the calculations and the experimental measurements for $D_{e}$.}
	\label{tab1}
	\begin{center}
	\begin{tabular}{lllcccc}
		\hline 
  Source & Level of & Basis & $R_{e}$ & $E(R_{e})$ & $D_{e}$ & $\epsilon_{0}$ \\
	    & theory & set & $(a_0)$ & (Hartree) & (eV) & $(\%)$ \\
     \hline
    Exp \cite{ram1985, hotop1998} & & & 1.8731 & - & 2.275 & \\
    Present & MRCI$^a$ & sec.~\ref{sec:computational} & 1.8920 & -128.8965 & 2.269 & 0.3\\
    Theory \cite{montes2021} & CCSD(T) & CBS & 1.8733 & -128.9536 & 2.292 & 0.7\\
    Theory \cite{montes2021} & MRCI+Q & AV6Z & 1.8733 & -128.9485 & 2.296 & 0.9\\
    Theory \cite{coxon2016} & - & - & 1.8729 & - & 2.179 & 4.4\\
    Theory \cite{gerivani2015} & MRCI & aug-AO & 1.8756 & - & 2.288 & 0.5\\
    Theory \cite{ngassam2008} & CI & TZP & 1.8935 & - & 2.320 & 1.9\\
    Theory \cite{lo2005} & SCF & cc-pVTZ & 1.8627 & -128.881 & 2.370 & 4.0\\
    Theory \cite{mitchell2005} & CI & aug-AO & 1.8708 & - & 2.148 & 5.9\\
    Theory \cite{civis2004} & MRCISD$^b$ & - & 1.8731 & - & 2.296 & 0.9\\
    Theory \cite{petsalakis1998} & MRDCI & aug-AO & 1.8948 & -128.9787 & 2.290 & 0.7\\
    Theory \cite{kraemer1997} & MRCI$^b$  & aug-AO & 1.8765 & - & 2.310 & 1.5\\
    Theory \cite{rosmus1980} & CEPA$^c$ & AO & 1.8820 & -128.3490 & 2.276 & 0.1\\
    Theory \cite{bondybey1972} & SCF & STO & 1.850 & -128.6242 & 2.10 & 8.3\\
    Theory \cite{bondybey1972} & CI & STO & 1.8689 & -128.6818 & 2.270 & 0.2\\
    Theory \cite{wan2019} & MRDCI  & aug-AO & 1.8670 & - & 2.430 & 6.4\\
    Theory \cite{yan2024} & MRCI  & cc-pV5Z & 1.8730 & - & 2.299 & 1.0\\
		\hline 
	\end{tabular}\\
\end{center}
 $^a$ Atomic orbitals calculated with MCSCF. \\
 $^b$ Atomic orbitals calculated with SCF-CASSCF. \\
 $^c$ Atomic orbitals calculated with SCEP. \\
 All other: Atomic orbitals calculated with SCF. \\
\end{table}

In order to fully characterize the molecular system, we have calculated the major spectroscopic data of the molecular target and for the neutral system. They are shown in 
tables~\ref{tab1},~\ref{tab2},~\ref{tab3} and \ref{tab4}, where our results are compared with previous theoretical results~\cite{mitchell2005,ngassam2008,petsalakis1998,lo2005, montes2021,coxon2016,gerivani2015,civis2004,kraemer1997,rosmus1980,bondybey1972,wan2019,yan2024}, experimental results~\cite{ram1985, hotop1998,civis2004}, and data from the NIST chemistry webbook~\cite{linstorm1998}.

Table~\ref{tab1} refers to the molecular ion, and shows values for its equilibrium geometry, the potential energy at equilibrium, and its dissociation energy. We compare our calculated parameters with other theoretical results~\cite{mitchell2005,ngassam2008,petsalakis1998,lo2005, montes2021,coxon2016,gerivani2015,civis2004,kraemer1997,rosmus1980,bondybey1972,wan2019,yan2024}, and evaluate relative differences of the dissociation energy in contrast to the experimental measurements~\cite{ram1985, hotop1998}. Our results agree well with existing experimental and other theoretical results, typically within 2\%, except for a few studies \cite{coxon2016,lo2005,mitchell2005,bondybey1972,wan2019} which employed lower levels of theory.

\begin{table}[t]
\caption{Calculated vibrational energy levels $E_{v_{i}}^{+}$ (eV) for NeH$^{+}$ ($X^{1}\Sigma^{+}$) with respect to $E_0 = -128.88998$ Hartree, and comparison with experimental values from~\cite{civis2004, ram1985}. The relative differences between our results and the experimental data are given by $\epsilon_{0}(\%)$.}
	\label{tab2}
	\centering
	\begin{tabular}{lcccccc}
		\hline
  \rule{0pt}{3ex} 
	$v_{i}$$^{+}$ & This work & Exp & $\epsilon_{0}$ (\%)\\
	    \hline
     \rule{0pt}{3ex}1 & 0.33524 & 0.33197 & 0.98\\
      2 & 0.64055 & 0.63593 & 0.72\\
      3 & 0.91920 & 0.91267 & 0.71\\
      4 & 1.16981 & 1.16138 & 0.72\\
      5 & 1.39213 & 1.38233 & 0.70\\
      6 & 1.58560 & 1.57472 & 0.69\\
      7 & 1.74969 & 1.73799 & 0.67\\
      8 & 1.88330 & 1.87132 & 0.64\\
      9 & 1.98643 & 1.97391 & 0.63\\
      10 & 2.05881 & 2.04629 & 0.61\\
      11 & 2.10262 & 2.08983 & 0.61\\
      12 & 2.12384 & 2.11051 & 0.63\\
      13 & 2.12384 & 2.11731 & 0.31\\
      14 & 2.12384 & 2.11895 & 0.23\\
	\hline
 \end{tabular}
\end{table}

In table~\ref{tab2}, we list all the 15 vibrational levels of the cation calculated using the Numerov-Cooley method and compare them with experimental observations~\cite{civis2004,ram1985}. The relative differences obtained are below 1\%, which confirms the accuracy of our structure calculations.

\begin{table}[t]
	\caption{
    Comparison of our computed NeH$^{+}$ (X $^{1}\Sigma^{+}$) and NeH Rydberg states dissociation limits relatively to Ne+H$(n=1)$~(eV) with previous theoretical results~\cite{petsalakis1998,lo2005} and data from NIST chemistry webbook ~\cite{linstorm1998}. The relative differences between our results and the NIST data are given by $\epsilon_{0}(\%)$.}
	\label{tab3}
	\centering
	\begin{tabular}{lccccccc}
		\hline
	level & This work & \cite{petsalakis1998} & \cite{lo2005} & NIST & $\epsilon_{0}$ (\%)\\
	    \hline
     \rule{0pt}{3ex}$n=2$ & 10.20509 & 10.21788 &  10.20209 & 10.19880 & 0.062\\
     $n=3$ & 12.09383 & 12.10853 & 12.09201 & 12.08750 & 0.052\\
     H IP & 13.55481 & 13.56815 &  13.57945 & 13.59840 & 0.322\\
     \hline
     	\end{tabular}
\end{table}

Table~\ref{tab3} gives the dissociation limits of the electronic states of  NeH$^{+}$ (X $^{1}\Sigma^{+}$) and of the NeH Rydberg states relative to Ne+H($n=1$), in comparison with previously computed limits~\cite{petsalakis1998,lo2005} and with data from the NIST chemistry webbook ~\cite{linstorm1998}.

\begin{table}[t]
	\caption{     
    Energy differences $\Delta$E(R$_{e}$) (eV) at the equilibrium distance R$_{e}$ of the NeH$^+$ ground electronic state, between several states and the NeH lowest excited electronic state (A $^{2}\Sigma^{+}$). The results are compared with the previous theoretically obtained  data~\cite{petsalakis1998,lo2005}.}
	\label{tab4}
	\centering
	\begin{tabular}{cccccccc}
     \hline
	State & This work & \cite{petsalakis1998} & \cite{lo2005}\\
	    \hline
     \rule{0pt}{3ex}B $^{2}\Pi$ & 0.12045 & 0.22365 &  0.17499 &\\
     C $^{2}\Sigma^{+}$ & 0.67190 & 0.70451 &  0.64747 &\\
     2 $^{2}\Pi$ & 1.69663 & 1.63268 &  1.66243 &\\
     4 $^{2}\Sigma^{+}$ & 1.70531 & 1.72214 &  1.62743 &\\
     5 $^{2}\Sigma^{+}$ & 2.08265 & 1.99053 &  - &\\
     X $^{1}\Sigma^{+}$ & 3.05459 & 3.14235 &  3.07987 &\\
        \hline
        \end{tabular}
\end{table}

Finally, table~\ref{tab4} contains the excitation energies of NeH$^+$ and of NeH electronic states relative to the A $^{2}\Sigma^{+}$ state of NeH for all electronic states above the same A $^{2}\Sigma^{+}$ state, calculated at the equilibrium distance of the ion. 

The corresponding energy differences $\Delta$E(R$_{e}$) are compared with the previous calculations~\cite{petsalakis1998,lo2005}. One may notice that, except for the B state, where a significant gap occurs, the agreement is satisfactory.

Our computational framework does not take into account the Rydberg states of NeH built on the excited states of the NeH$^+$ ion. For several molecular species, these 'intruder' states play a visible role in the spectroscopy of the neutral and in its dynamics. However, this happens only when the lowest excited states of the ion are close to the ground one, which is not the case for NeH$^+$~\cite{gerivani2015}. Therefore, we believe that our results are accurate, since not being affected by the core-excited Rydberg states effects.

\section{Conclusions}{\label{sec:conclusions}}

The potential energy curves for the ground state $^{1}\Sigma^{+}$ of NeH$^{+}$ ion and for the lowest five $^{2}\Sigma^{+}$ states and two $^{2}\Pi$ states of NeH molecule have been computed with high precision using the MOLPRO quantum chemistry suite at the MCSCF-MRCI level of theory, employing an extended complete active space (CAS) and augmented basis sets optimized for dissociation up to Ne+H($n=3$). The results show good agreement with previous experimental and theoretical studies, particularly with the work of Theodorakopoulos et al.~\cite{Theodorakopoulos1987}, Petsalakis et al.~\cite{petsalakis1998}, and Lo et al.~\cite{lo2005} for some of these states. While the non-adiabatic couplings $A(R)$ between the states (A-C, X-A, X-C) up to Ne+H($n=2$) had already been addressed by Theodorakopoulos et al.~\cite{Theodorakopoulos1987}, our calculations went beyond this producing further, namely both the couplings $A(R)$ and $B(R)$ between states dissociating up to the limit Ne+H($n=3$) -  X-4, A-4, C-4, X-5, A-5, C-5 for the $^{2}\Sigma^{+}$ symmetry, and B-2 for the $^{2}\Pi$ symmetry. These couplings are essential for the calculation of cross sections and rate coefficients in low-energy collisions between electrons and NeH$^{+}$~\cite{Giusti1980, Mezei2019}. Work on this is currently in progress. 

\section*{Data availability}
The data underlying this article will be shared on reasonable request to the corresponding author.

\begin{acknowledgments}
The authors acknowledge support provided by the F\'ed\'eration de Recherche Fusion par Confinement Magn\'etique (CNRS and CEA), La R\'egion Normandie, LabEx EMC3 through projects PTOLEMEE,  COMUE Normandie Universit\'e and the Institute for Energy, Propulsion and Environment (FR-IEPE). 
J.Zs.M. is grateful for financial support from the National Research, Development and Innovation Fund of Hungary, under the FK 19 funding schemes with project number FK 132989.  
This work was granted access to the HPC/AI resources of [CINES/IDRIS/TGCC] under the allocation 2023-2024 [AD010805116R2] made by GENCI.
\end{acknowledgments}

%
\bibliographystyle{unsrt}
\bibliography{aipsamp}

\end{document}